\documentclass[a4paper,twocolumn,10pt]{article}

\usepackage[margin=1in]{geometry}

\pdfoutput=1

\usepackage{graphicx}
\usepackage{bm}
\usepackage{dcolumn}
\usepackage{setspace}
\usepackage{abstract}
\usepackage{multirow}
\usepackage{sidecap}
\usepackage{caption}
\usepackage{subcaption}

\usepackage[affil-it]{authblk}

\usepackage[utf8]{inputenc}
\usepackage{lineno,hyperref}

\begin{document}
\title{ \bf Directional Anisotropy of Crack Propagation Along $\Sigma$3 Grain Boundary in BCC Fe}

 \date{}

\author{ G. Sainath*, B.K. Choudhary** }

\affil {Deformation and Damage Modeling Section, Mechanical Metallurgy Division \\ 
Indira Gandhi Centre for Atomic Research, Kalpakkam - 603102, Tamil Nadu, India}
\onecolumn

\maketitle
\doublespacing
\begin{onecolabstract}

Crack growth behaviour along the coherent twin boundary (CTB), i.e., $\Sigma$3\{112\} of BCC Fe is investigated 
using molecular dynamics (MD) simulations. The growth of an atomistically sharp crack with \{112\}$<$110$>$ 
orientation has been examined along the two opposite $<$111$>$ directions of CTB under mode-I loading at a
constant strain rate. Separate MD simulations were carried out with crack inserted in the left side, right 
side and middle of the specimen model system. The results indicate that the crack grows differently along 
the two opposite $<$111$>$ directions. In case of a crack inserted in the left side, the crack grows in 
ductile manner, while it propagates in semi-brittle manner in the case of crack inserted in the right side. 
The directional dependence of crack growth along the CTB is also confirmed by the stress-strain behaviour. 
This anisotropy in crack growth behaviour has been attributed to the twinning-antitwinning asymmetry of 
1/6$<$111$>$ partial dislocations on \{112\} planes. \\

\noindent {\bf Keywords: } Molecular dynamics simulations, Crack propagation, $\Sigma$3 grain boundary, ductile 
and brittle


\end{onecolabstract}

\renewcommand{\thefootnote}{\fnsymbol{footnote}} \footnotetext{* email : sg@igcar.gov.in}
\renewcommand{\thefootnote}{\fnsymbol{footnote}} \footnotetext{** e-mail : bkc@igcar.gov.in}

{\small 

\section{Introduction}

The properties of a polycrystalline material are controlled by the grain boundaries (GBs). The GBs serve as 
preferential sites for the nucleation of dislocations and cracks, and can act as preferred path for crack 
propagation. It can also impede the motion of dislocations and the growth of cracks. The crack propagation 
along the GBs is known as intergranular fracture. The intergranular fracture resulting from the nucleation 
and growth of wedge cracks and cavities is an important mode of failure in creeping solids. In view of this, 
understanding the crack growth behaviour along the GBs becomes important in order to design the creep resistant 
microstructure. Earlier studies on GB cracking in bicrystals have shown that the brittle or, ductile response 
of GBs depends on the direction of crack advance \cite{Wang,Wang-2}. However, introducing the crack along the 
particular GB and studying its propagation using experimental techniques is difficult and challenging. In this 
context, atomistic simulations can be effectively used to understand the crack propagation behaviour along the 
particular grain boundary.

Molecular dynamics (MD) simulations have been widely used to examine crack propagation in solids. Cheng et al. 
\cite{Cheng} examined intergranular fracture along the different GBs in Cu using MD simulations. Along the 
$\Sigma 3(1\bar11)$ and $\Sigma 11(1\bar13)$ coherent GBs, directional anisotropy in terms of brittle cleavage 
in one direction and dislocation emission in the opposite direction has been observed \cite{Cheng}. However, 
this directional anisotropy is not observed along the $\Sigma 11(3\bar32)$ and $\Sigma 9(2\bar21)$ incoherent 
GBs, where the crack propagates in a ductile manner in both the directions \cite{Cheng}. The directional 
anisotropy in crack propagation has also been observed for $\Sigma$ 29(520) GBs in NiAl \cite{NiAl} and 
$\Sigma 29 (5\bar 57)$ GBs in Al \cite{Al}. These investigations clearly suggest that the nature of GBs play 
an important role during crack propagation. A large scale MD simulations project has been undertaken at IGCAR to 
investigate the nucleation and growth of damage along the different coincidence site lattice (CSL) boundaries in 
BCC Fe. These simulations are intended to effectively supplement the ongoing research activities on the 
improvement of high temperature mechanical properties of reactor structural and steam generator materials 
using grain boundary engineering route. In the present paper, we examine the crack propagation behaviour 
along the $\Sigma$ 3(112) GBs in BCC Fe in the positive and negative  $<$111$>$ directions. The $\Sigma$ 3(112) 
GB in BCC Fe is a coherent twin boundary (CTB) and is the simplest of all the boundaries, in which one grain 
is a mirror reflection of the other.

\section{Simulation Details}

Molecular dynamics simulations have been performed using Large scale Atomic/Molecular Massively Parallel 
Simulator (LAMMPS) package \cite{LAMMPS} employing an embedded atom method potential for BCC Fe given by 
Mendelev and coworkers \cite{Mendelev}. The Mendelev potential is widely used to study the deformation 
behaviour of BCC Fe \cite{Sainath-1,Sainath-2,Sainath-orient}. The visualization of atomic snapshots is 
accomplished in AtomEye \cite{Atomeye} using centro-symmetry parameter (CSP) coloring \cite{CSP}. Initially, 
single crystal BCC Fe oriented in [110], $[1\bar11]$ and $[1\bar1\bar2]$ crystallographic directions was 
chosen (Fig. \ref{Initial}). This specimen model had the dimensions of 1.4 $\times$ 17.3 $\times$ 17.3 nm 
containing about 42,000 atoms. In this specimen model, the CTBs were introduced by $180^o$ rotation about 
the $[1\bar1\bar2]$ axis and the model represents a multilayer system with each layer of Fe separated by 
CTB. The atomistically sharp crack was introduced on CTB with the crack front being the [110] direction 
(Fig. \ref{Initial}). The crack length was half the width of the specimen. Periodic boundary conditions 
were applied only in the crack front direction. In order to study the crack growth along the mutually 
opposite directions of CTB, the MD simulations were carried out at 10 and 600 K with crack being inserted 
in the left side (Fig. \ref{Initial}a), right side (Fig. \ref{Initial}b) and middle of the specimen 
(Fig. \ref{Initial}c). Here onwards, these models have been referred to as left, right and middle models. 
The model systems were equilibrated to the respective desired temperatures in NVT ensemble. Upon completion 
of equilibrium process, the growth of an atomically sharp crack has been studied along CTB under mode-I 
loading at a constant strain rate of 1 $\times 10^8$ $s^{-1}$. The fixed strain rate is achieved by imposing 
velocity to atoms along the $[1\bar1\bar2]$ axis that varied linearly from zero at the bottom to a maximum 
value at the top layer. The average stress is calculated from the Virial expression \cite{Virial}.

\begin{figure}[h]
\centering
\includegraphics[width=12cm]{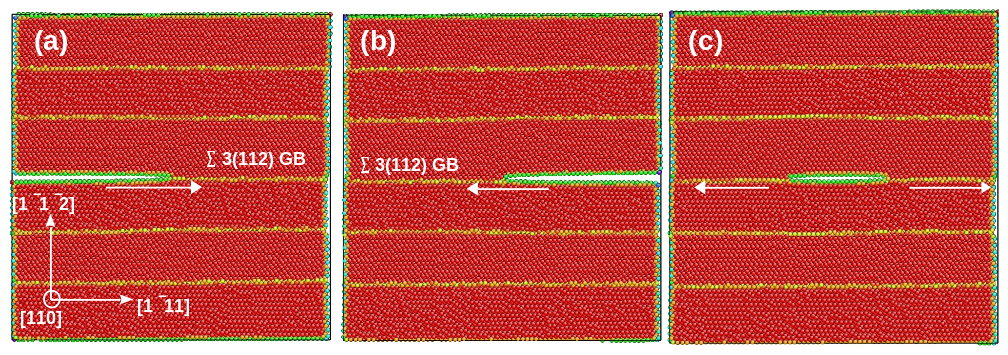}
\caption {\footnotesize The initial specimens showing atomistically sharp crack located at the (a) left side (b) right 
side and (c) middle on the GBs of the model system. The atoms are colored according to CSP \cite{CSP}. The white 
arrows indicate the direction of crack propagation.}
\label{Initial}
\end{figure}

\section{Results and Discussion}

The atomic snapshots depicting the crack propagation behaviour from left to right along the CTB of BCC Fe at
10 K are shown in Fig. \ref{Left}. This direction of crack propagation from left to right has been referred 
to as positive $<$111$>$ direction. It can be seen that the yielding occurs by the nucleation of a twin embryo 
consisting of many 1/6$<$111$>$ partial dislocations \cite{Sainath-orient} ahead of crack tip from the GB 
(Fig. \ref{Left}a). With increasing strain, the twin embryo reaches the next or, 4th CTB and becomes full 
twin as shown in Fig. \ref{Left}b. This process of twin formation creates a step on the 4th CTB, while it 
annihilates the part of the 3rd CTB on which it is nucleated (Fig. \ref{Left}b). With further increase in 
strain, more twin embryos/twins nucleate from CTB (Fig. \ref{Left}c), and as a result, the 3rd CTB becomes
curved in nature. The continuous nucleation of twin embryos and their interaction with 4th CTB leads to the
migration of 3rd and 4th CTBs (Fig. \ref{Left}d, e). Following the migration of 3rd CTB, the crack tip emits 
1/6$<$111$>$ partial dislocations and this results in crack blunting as shown in Fig. \ref{Left}e. Thus, the 
crack in this case propagates in ductile manner along with grain boundary migration.

\begin{figure}[h]
\centering
\includegraphics[width=10cm]{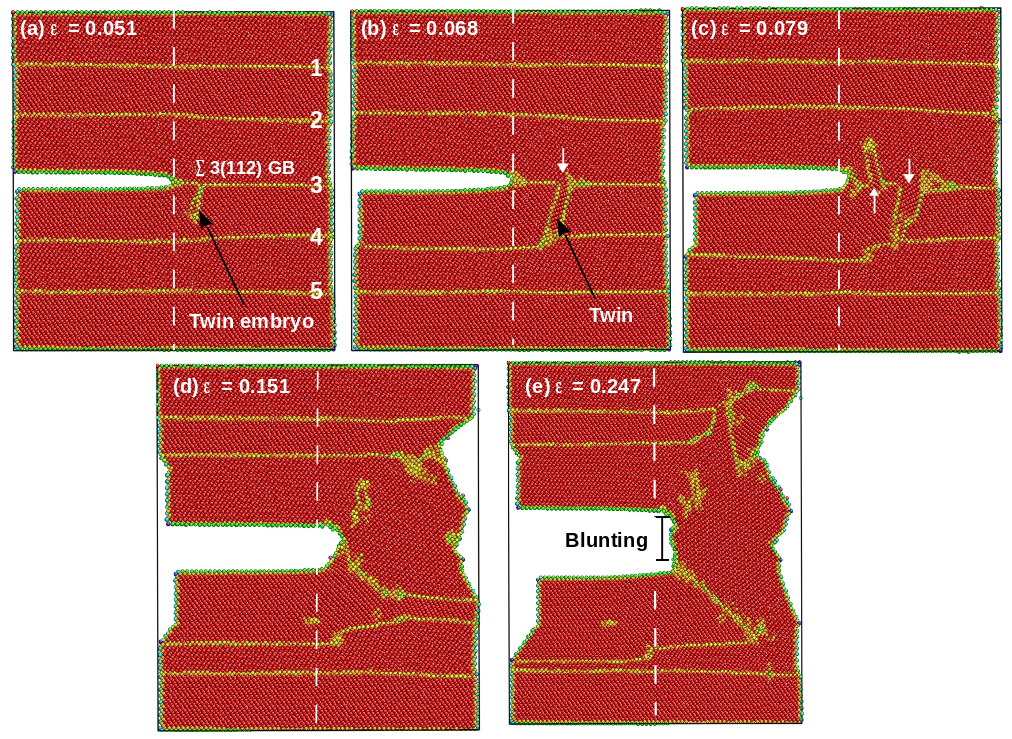}
\caption {\footnotesize The crack propagation behaviour from left to right along the CTB in BCC Fe. The dashed vertical 
white line shows the initial crack tip position.}
\label{Left}
\end{figure}

The atomic snapshots depicting the crack propagation behaviour from right to left along the CTB in BCC Fe at
10 K are shown in Fig. \ref{Right}. This direction of crack propagation from right to left has been referred 
to as negative $<$111$>$ direction. Contrary to crack propagating from left to right (Fig. \ref{Left}a), the twin 
embryo nucleates directly at the crack tip when the crack propagates from right to left (Fig. \ref{Right}a).
Further, the strain at which the twin embryo nucleates has been found to be considerably lower than the 
previous case. With increasing strain, the twin embryo reaches the next or 4th CTB and becomes full twin as 
shown in Fig. \ref{Right}b. This process of twin formation annihilates the part of 4th CTB (Fig. \ref{Right}b). 
With further increase in strain, more and more 1/6$<$111$>$ partial dislocations nucleates directly from the
crack tip and propagates along the twin boundaries of the newly formed twin. This leads to the rapid crack 
propagation along the 3rd CTB without considerable plastic deformation (Fig. \ref{Right}c-e). This suggests 
that the crack propagation in this case remains semi-brittle.

\begin{figure}[h]
\centering
\includegraphics[width=10cm]{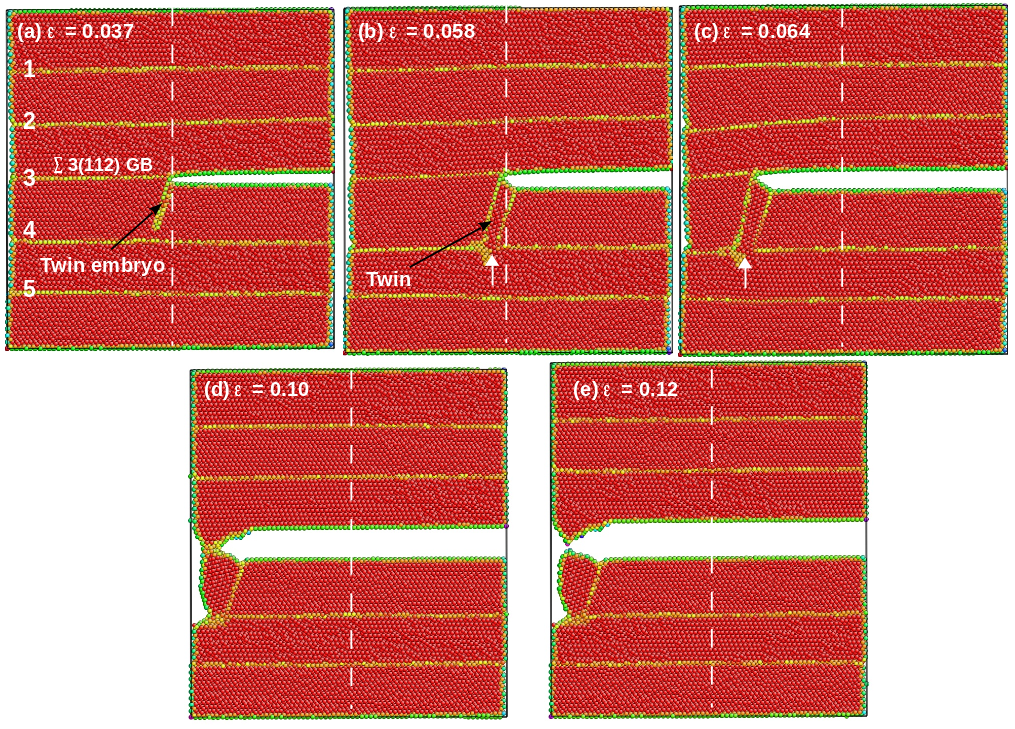}
\caption {\footnotesize The crack propagation behaviour from right to left along the CTB in BCC Fe. The dashed vertical 
white line shows the initial crack tip position.}
\label{Right}
\end{figure}

In order to confirm the above observed anisotropy, MD simulations performed on the specimen with the crack
inserted in the middle of CTB (Fig. \ref{Initial}c) are shown in Fig. \ref{Middle}. It can be seen that the 
crack propagates both the left and right directions simultaneously. When the crack propagates from the left 
to right, the nucleation of twin embryo from the GBs ahead of crack tip (Fig. \ref{Middle}b) and the twin 
boundary migration along with crack blunting (Fig. \ref{Middle}c-e) can be seen. On the other hand, in case 
of crack propagating from right to left, nucleation of twin embryo directly at the crack tip (Fig. \ref{Middle}a, b) 
and rapid crack propagation along the twin boundary of the newly formed twin (Fig. \ref{Middle}c-e) have been
observed. The stress-strain behaviour of BCC Fe containing the left and right crack along CTB (Fig. \ref{Initial}a, b) 
is shown in Fig. \ref{stress-strain}. Both the left and right models show similar stress-strain behaviour during 
elastic deformation. However, the right model yields at comparatively lower stress/strain and exhibits lower 
ultimate tensile strength (UTS) than those observed for the left model. The yielding by the twin embryos 
nucleation is followed by hardening due to twin-twin interactions up to UTS (Figs. \ref{Left}b, \ref{Right}b). 
Beyond UTS, the stress-strain behaviour of right and left models differs drastically: the stress in the right 
model abruptly drops to low value concurrent with rapid crack propagation and early failure by semi-brittle 
manner, while the stress in the left model fluctuates at higher values displaying ductile nature of crack 
propagation.

\begin{figure}[h]
\centering
\includegraphics[width=10cm]{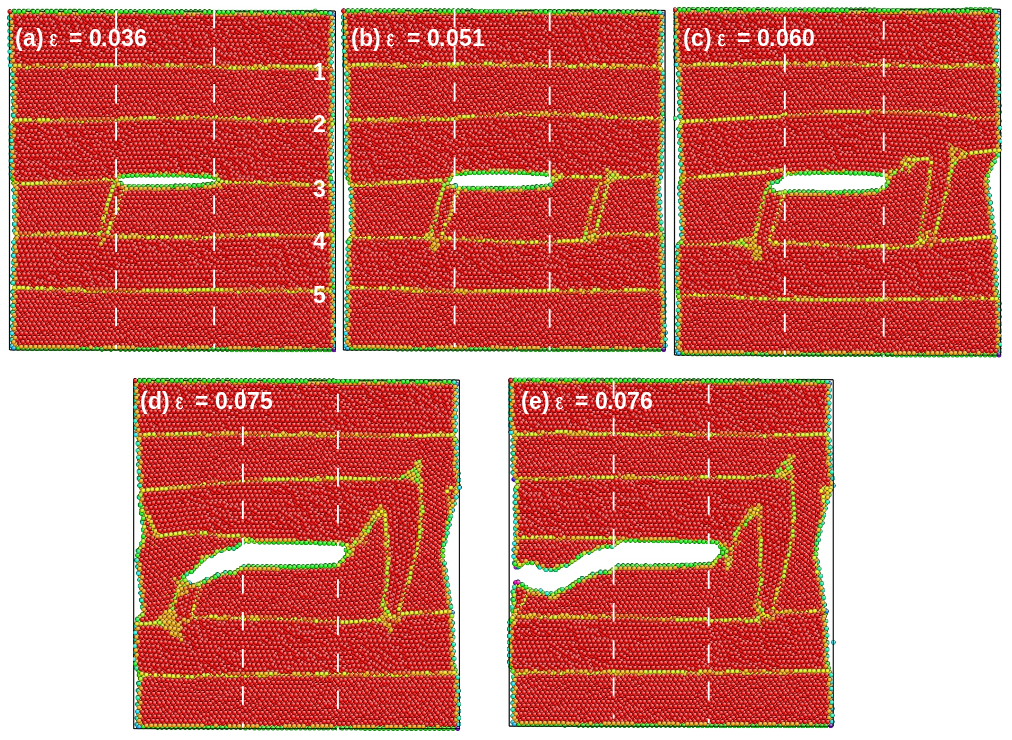}
\caption {\footnotesize The propagation in both positive and negative $<$111$>$ directions of a crack located in the middle 
of specimen. The dashed vertical white lines show the initial crack tip positions.}
\label{Middle}
\end{figure}

\begin{figure}[h]
\centering
\includegraphics[width=7cm]{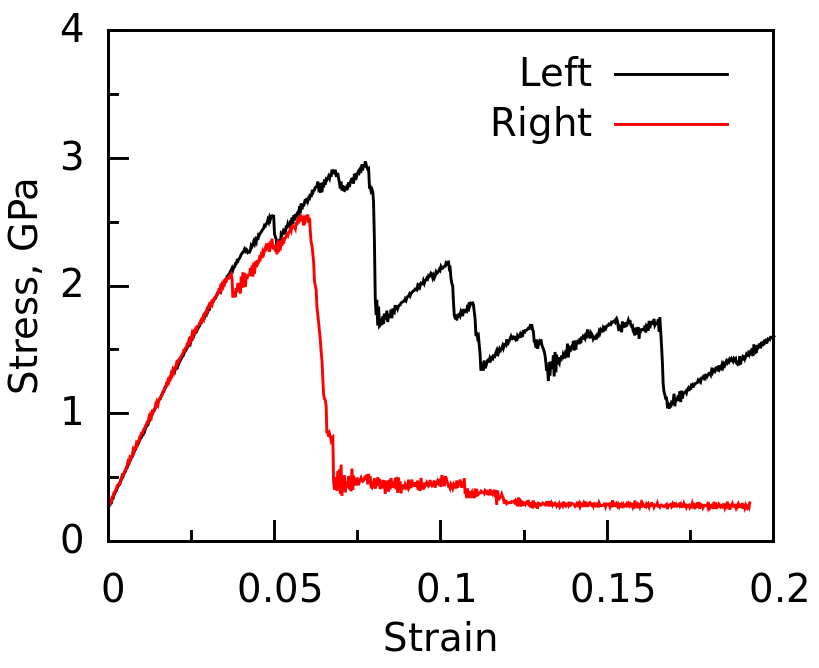}
\caption {\footnotesize The stress–strain behaviour of BCC Fe containing the left and right crack along $\Sigma$3(112) 
grain boundary.}
\label{stress-strain}
\end{figure}

The observed directional anisotropy of crack propagation is in agreement with that reported for the CTB 
($\Sigma3(111)$) in FCC Cu \cite{Cheng}. The directional anisotropy of crack propagation has been 
explained by the Rice model of ductile and brittle fracture \cite{Rice}. According to Rice model, the 
brittle or, ductile behaviour of a given GB can be understood by comparing the values of energy release 
rate associated with brittle cleavage ($G_{cleav}$) and the dislocation nucleation from a crack tip ($G_{disl}$) 
for all possible slip systems \cite{Cheng, Kulkarni}. When $G_{cleav} < G_{disl}$, the GB behaves in a brittle 
manner, and for the case of $G_{cleav} > G_{disl}$, the GB exhibits ductile behavior. The directional anisotropy of 
crack propagation along the CTB observed in the present investigation in BCC Fe can be attributed to the 
twinning-antitwinning asymmetry of 1/6$<$111$>$ partial dislocations along the \{112\} planes \cite{asymmetry}. 
The glide of 1/6$<$111$>$ partial dislocations on \{112\} planes is allowed only in one direction (twinning sense) 
and the glide in opposite direction (antitwinning direction) creates an unstable stacking fault \cite{Healy}. 
When the crack propagates from left to right (Fig. \ref{Left}), the 1/6$<$111$>$ partial dislocations glide 
in a twinning sense on 3rd CTB and this leads to crack blunting and twin boundary migration. On the other hand, when the 
crack propagates from right to left (Fig. \ref{Right}), the 1/6$<$111$>$ partial dislocations will have an 
antitwinning sense on 3rd CTB (the 1/6$<$111$>$ partials have high critical resolved shear stress in antitwinning 
sense), and this leads to twin embryo nucleation on the \{112\} plane other than the 3rd CTB. The rapid migration 
of this twin boundary mediates the crack propagation leading to brittle failure. At higher temperature of 600 K, 
similar directional anisotropy of crack propagation in BCC Fe has been observed.

\section{Conclusions}
The crack propagation behaviour along the CTB in BCC Fe has been studied along the opposite $<$111$>$ directions 
using MD simulations. The simulation results show that the crack propagation along the CTB exhibits directional 
anisotropy. In one direction, it propagates in a ductile manner accompanied by twin boundary migration, while in 
the opposite direction the crack propagates rapidly along the CTB in a semi-brittle manner. Similar results were
observed when the crack is allowed to propagate simultaneously in these two directions. The directional anisotropy
of crack propagation is also reflected in the stress-strain behaviour. This anisotropy in crack growth behaviour 
has been attributed to the twinning-antitwinning asymmetry of 1/6$<$111$>$ partial dislocations on \{112\} planes 
in BCC Fe. 

}

\end{document}